\documentclass[peerreview]{IEEEtran}

\usepackage[utf8]{inputenc} 
\usepackage[T1]{fontenc}
\usepackage{url}
\usepackage{ifthen}
\usepackage{cite}
\usepackage[cmex10]{amsmath} 

\usepackage{graphicx}
\usepackage{amsmath,amsthm,amsfonts,mathtools}
\usepackage{dsfont}

\usepackage{algorithm}
\usepackage{algpseudocode}

\usepackage{color}
\usepackage{float}

\newcommand{\rem}[1]{}

\newtheorem{theorem}{Theorem}

\newtheorem{definition}{Definition}

\newtheorem{example}{Example}

\newcommand{\bre}{\begin{equation}}
\newcommand{\ere}{\end{equation}}

\newcommand{\ee}\]
\newcommand{\bra}{\begin{eqnarray}}
\newcommand{\era}{\end{eqnarray}}
\newcommand{\bfg}{\begin{figure}[hbtp]}
\newcommand{\efg}{\end{figure}}
\newcommand{\bver}{\begin{verbatim}}
\newcommand{\ever}{\end{verbatim}}
\newcommand{\bit}{\begin{itemize}}
\newcommand{\eit}{\end{itemize}}
\newcommand{\ben}{\begin{enumerate}}
\newcommand{\een}{\end{enumerate}}

\newcommand{\bc}{{\bf c}}

\newcommand{\talpha}{\tilde {\alpha}}

\newcommand{\GF}{{\mathrm {GF}}}

\newcommand{\baa}{\begin{eqnarray*}}
\newcommand{\eaa}{\end{eqnarray*}}

\newcommand{\bv}{{\bf v}}

\newcommand{\bw}{{\bf w}}

\newcommand{\cB}{{\cal B}}

\newcommand{\cV}{{\cal V}}
\newcommand{\cJ}{{\cal J}}

\newcommand{\cC}{{\mathcal{C}}}

\newcommand{\cN}{{\mathcal{N}}}
\newcommand{\cG}{\mathcal{G}}

\newcommand{\beginproof}{\noindent \textbf{Proof: }  }
\newcommand{\finproof}{\noindent $\Box$\\}
\newcommand{\defined}{\triangleq}

\def\defined{\: {\stackrel{\scriptscriptstyle \Delta}{=}} \: }

\newfont{\boldlarge}{msbm10 scaled 1100}

\newlength{\tmpbigbar}

\begin{document}
\title{On the error correction of iterative bounded distance decoding of generalized LDPC codes}

\author{David Burshtein
	\thanks{D.\ Burshtein is with the School of Electrical and Computer Engineering, Tel-Aviv University, Tel-Aviv 6997801, Israel (email: burstyn@eng.tau.ac.il).}
}

\maketitle \setcounter{page}{1}

\begin{abstract}
Consider an ensemble of regular generalized LDPC (GLDPC) codes and assume that the same component code is associated with each parity check node. To decode a GLDPC code from the ensemble, we use the bit flipping bounded distance decoding algorithm, which is an extension of the bit flipping algorithm for LDPC codes.
Previous work has shown conditions, under which, for a typical code in the ensemble with blocklength sufficiently large, a positive constant fraction of worst case errors can be corrected.
In this work we first show that these requirements can be relaxed for ensembles with small left degrees.
While previous work on GLDPC codes has considered expander graph arguments, our analysis formulates a necessary condition that the Tanner graph needs to satisfy for a failure event and then shows that the probability of this event vanishes for a sufficiently large blocklength.
We then extend the analysis to random error correction and derive a lower bound on the fraction of random errors that can be corrected asymptotically. We discuss the extension of our results to non-binary GLDPC codes and present numerical examples.
\end{abstract}

\section{Introduction} \label{sec:introduction}
Following the introduction of low-density parity-check (LDPC) codes by Gallager \cite{galmono}, Tanner \cite{tanner} has proposed generalized LDPC (GLDPC) codes. In a GLDPC code, each parity check node in the Tanner graph is associated with a general linear component code which is not necessarily a single parity check code as in LDPC codes.
The properties of GLDPC codes under maximum likelihood (ML) and under iterative decoding, as well as their actual performance for various code architectures and communication channels, were studied in \cite{boutros1999generalized,lent_zig_GLDPC,liva2008quasi,miladinovic2008generalized,paolini2010generalized,flanagan2011growth,liu2018generalized,liu2019probabilistic,farooq2020generalized,liu2020design,mitchell2021spatially,rezgui2021class,bocharova2022irregular} and references therein.
Due to the generalized structure, iteratively decoded GLDPC codes can reduce the error rate compared to plain LDPC codes.
In addition, although the computational complexity of a single iterative decoding iteration is higher, GLDPC codes can reduce the total required number of iterations.

In this paper we consider the ability of GLDPC codes to correct worst case (adversarial) errors.
An ensemble of (e.g., LDPC or GLDPC) codes can correct a (positive) constant fraction of worst case errors, using low complexity iterative decoding, if there exists $\alpha>0$ such that, for blocklength, $N$, sufficiently large, a typical code from the ensemble can correct all possible patterns of at most $\alpha N$ errors in any transmitted codeword. That is, the error correction radius of almost all codes in the ensemble is at least $\alpha N$ so that it scales linearly with the blocklength.

Some authors have analyzed the worst case error correction of LDPC codes. Zyablov and Pinsker \cite{zyablov} have shown that $(c\ge 5,d)$-regular LDPC codes can correct a constant fraction of worst case errors when using the iterative low complexity bit flipping algorithm to decode. 
Sipser and Spielman \cite{sipser} have analyzed the error correction capability of expander codes, decoded using the bit flipping algorithm. Since a typical code in the $(c\ge 5,d)$-regular ensemble of LDPC codes, with $N$ sufficiently large, has the expansion required in \cite{sipser}, one consequence of this work is an alternative proof to the above mentioned result in \cite{zyablov}.
We note that a different line of research examined expander codes which are special cases of (doubly) generalized LDPC codes with a sufficiently large blocklength component code \cite{sipser,zemor2001expander,barg2002error,guruswami2002near,roth2006improved}. 
In \cite{lp_linear_err_correct1_journal} it was shown that the linear programming decoder can also correct a constant fraction of worst case errors in an LDPC code with $N$ sufficiently large that has the required expansion.
In \cite{bur_expander} it was shown that expander graph arguments also apply to message passing decoding algorithms, so that we do not really need to use the bit flipping algorithm to decode.
In \cite{bur_flipping}, it was shown that the condition for correcting a constant fraction of worst case errors can be relaxed to $c\ge 4$. However, for $c=4$, the bit flipping decoding needs to be modified to a probabilistic algorithm, and the error correction property (for a typical code in the ensemble) now holds for a typical sequence of random bits used by the probabilistic algorithm.
One consequence of these results is that, for a typical code drawn from the $(c\ge 4,d)$-regular ensemble, if we start the decoding with the belief propagation algorithm and then possibly switch to bit flipping decoding in the final iterations, iterative decoding of the code will correct, with probability that approaches one as $N\rightarrow\infty$, any fraction of random errors up to the asymptotic belief propagation threshold of the ensemble.
A different approach was used in \cite{lentmaier2005analysis}. For $(c\ge 3,d)$-regular LDPC codes with $N$ sufficiently large, it was shown that there exist codes with block error rate that approaches zero when we operate below the asymptotic belief propagation threshold. This result is also generalized to GLDPC and Turbo codes. However, it does not apply to a typical code in the ensemble and \cite{lentmaier2005analysis} does not consider worst case error correction.
The work in \cite{zyablov} was extended to GLDPC codes in \cite{zyablov2009low}.
In \cite{chilappagari2010trapping} it was shown that a GLDPC code, with $N$ sufficiently large and a Tanner graph with sufficient expansion, can correct a constant fraction of worst case errors using the parallel bit flipping bounded distance decoding algorithm, which is an extension of the bit flipping algorithm for LDPC codes, and can be simulated sequentially in linear time. In addition to that, the paper establishes bounds on the number of worst case errors that can be corrected, as a function of the girth of the Tanner graph as  well as other parameters, both for LDPC and GLDPC codes.

Consider an ensemble of $(c,d)$-regular GLDPC codes and assume that the component code associated with each parity check node has minimum distance at least $2t+1$ for some $t\ge 1$. The codes are decoded using the bit flipping bounded distance decoding algorithm.
As explained below, the condition in \cite{chilappagari2010trapping} required for showing that the $(c,d)$-regular ensemble code can correct a constant fraction of worst case errors for $N$ sufficiently large is $c=3$ and $t\ge 3$, or $c=4$ and $t \ge 2$, or $c\ge 5$ and $t\ge 1$.
In this work we first show that these requirements can be relaxed as follows: For $c=3$ it is sufficient to require $t\ge 2$, and for $c=4$ it is sufficient to require $t\ge 1$.
Small values of $c$ are important since they are used in practical implementations \cite{liu2018generalized,liu2020design}.
As we note below, the recent work \cite{cheng2024can}, which uses expander graph arguments, implies that for $c=3$ it is sufficient to require $t\ge 2$ in order to decode a constant fraction of worst case errors in linear time. However, this result requires a much more complicated decoding algorithm with a much higher decoding time. In fact, the decoder in \cite{cheng2024can} activates the parallel bit flipping bounded distance decoding algorithm for a large (but constant) number of times, under an appropriate control logic. On the other hand, our result only uses a single run of the simple parallel bit flipping bounded distance decoding algorithm.
Our analysis is based on an extension of the technique we used in \cite{bur_flipping} to GLDPC codes.
While previous work has considered expander graph arguments \cite{chilappagari2010trapping}, our analysis formulates a necessary condition that the Tanner graph needs to satisfy for a failure event and then shows that the probability of this event vanishes for a sufficiently large blocklength, $N$.
We also extend the analysis to random error correction and derive a lower bound on the fraction of random errors that can be corrected asymptotically (large $N$). Then, we discuss the extension of our results to non-binary GLDPC codes and present numerical examples which demonstrate that our method also yields significantly larger values of the fraction of worst case errors, $\alpha$, that can be corrected compared to expander graph arguments. We also demonstrate a much larger fraction of random errors that can be corrected compared to the fraction of correctable worst case errors.

This paper is organized as follows. In Section \ref{sec:background} we provide background on GLDPC codes and recall some known bounds on multinomial coefficients. In Section \ref{sec:regular_GLDPC} we present our new results on the worst case error correction of regular binary GLDPC codes. In Section \ref{sec:extensions} we present extensions to random error correction of regular GLDPC codes and to non-binary regular GLDPC codes. Section \ref{sec:numerical_examples} provides some numerical examples. Section \ref{sec:conclusion} concludes the paper.

\section{Background} \label{sec:background}
\subsection{Generalized low-density parity-check (GLDPC) codes} \label{sec:background_GLDPC}
Consider a $(c, d)$-regular GLDPC bipartite graph-based code ensemble with blocklength $N$. This ensemble is constructed similarly to a plain $(c, d)$-regular LDPC bipartite graph-based code ensemble~\cite{luby2001improved},~\cite{urbanke_capacity}, except that now each check node does not necessarily represent a single parity code. Instead, the component code, $\cC_0$ (in the regular GLDPC codes considered in this paper, we assume the same component code for each check node), is a linear code with blocklength $d$ and rate $R_0=k_0/d$, where $k_0$ is the dimension of $\cC_0$. The minimum distance, $d_{\min}$, of $\cC_0$ is at least $2t+1$ for some $t>0$. Hence, with bounded distance decoding, one can correct any codeword of $\cC_0$ corrupted by $t$ errors or less.
Since $d\ge d_{\min} \ge 2t+1$,
\begin{equation}
\label{eq:d_t_ineq}
d > t+1 \: .
\end{equation}
Each code in the ensemble, defined in terms of its Tanner graph, has $N$ variable nodes, $\cV$, and $J$ check nodes, $\cJ$.
For each variable (check, respectively) node we assign $c$ ($d$) variable (check) sockets.
The total number of variable sockets is $Nc$ and the total number of check sockets is $J d$.
Hence, $Nc = Jd$.
We then choose a permutation with uniform probability from
the space of all permutations of size $N c$, and use this permutation to match the variable and check sockets.
Note that in this way multiple edges may link a pair of nodes.

Since each check node imposes $d-k_0$ constraints, the total number of independent parity check equations is at most $(d-k_0)J = (d-k_0) N c / d$, and the code rate, $R$, of each code in the ensemble is lower bounded by the nominal rate of the ensemble,
\begin{equation}
R \ge 1 - \frac{(d-k_0)c}{d} = 1 - (1-R_0)c \: .
\label{eq:R_GLDPC}
\end{equation}

Binary GLDPC codes are defined over $\GF(2)$, so that each variable node takes a value in $\{0,1\}$, and the component code, $\cC_0$ is binary.
Non-binary GLDPC codes are defined over a finite field, $\GF(q)$, with $q>2$, so that each variable node takes a value in $\{0,1,\ldots,q-1\}$, and $\cC_0$ is a code defined over $\GF(q)$.
In the general case, the rates $R_0$ and $R$ above are expressed in $q$-ary symbols (bits for $q=2$) per channel use.

\subsection{Bounds on multinomial coefficients} \label{sec:multinomial}
For non-negative integers $n_1,\ldots,n_i$ such that $\sum_{j=1}^i n_j \le n$, the multinomial coefficients are defined as,
\begin{equation}
\label{eq:choose}
\binom{n}{n_1,n_2,\ldots,n_i} \defined \frac{n!}{n_1! \ldots n_i! n_{i+1}!}
\end{equation}
where $n_{i+1} \defined n - \sum_{j=1}^i n_j$. Denote $\tau_j \defined n_j / n$, for $j=1,\ldots,i+1$ such that
$\sum_{j=1}^{i+1} \tau_j = 1$.
It can be shown (see e.g., \cite[Appendix I]{bur_flipping}) that,
\begin{equation}
\label{eq:multi_bnd_1}
\binom{n}{n_1,n_2,\ldots,n_i}
\le
e^{n h(\tau_1,\ldots,\tau_i)}
\end{equation}
where
\begin{equation}
\label{eq:entropy}
h(\tau_1,\ldots,\tau_i) \defined -\sum_{j=1}^i \tau_j \log \tau_j -
\left( 1-\sum_{j=1}^i \tau_j \right) \log \left( 1-\sum_{j=1}^i \tau_j \right)
\end{equation}
is the entropy function.
Note that the base of all the logarithms in this paper is $e$.
This bound can be improved by Stirling's formula \cite[Appendix I]{bur_flipping} under the assumption $n_j\ge 1$, $j=1,\ldots,i+1$:
\begin{equation}
\label{eq:multi_bnd_2}
\binom{n}{n_1,n_2,\ldots,n_i}
<
\left( 2\pi \right)^{-i/2} e^{1/12} \cdot 
\sqrt{\frac{n}{n_1 \cdot n_2 \cdot \ldots \cdot n_i \cdot \left( n-\sum_{j=1}^{i} n_j \right)}} \cdot
e^{nh(\tau_1,\ldots,\tau_i)}
\: .
\end{equation}
Note: If for some $j$, $n_j=0$, then we can remove it from the multinomial coefficient, decrease $i$ by 1, and then apply \eqref{eq:multi_bnd_2}.

Furthermore, by Stirling's formula, for $n_j \ge 1$, $j=1,\ldots,i+1$,
\begin{equation}
\label{eq:multi_upbnd}
\binom{n}{n_1,n_2,\ldots,n_i}
>
C_0 \cdot \frac{e^{nh(\tau_1,\ldots,\tau_i)}}
{\sqrt{n_1 \cdot n_2 \cdot \ldots \cdot n_i}}
\end{equation}
for
\begin{equation}
C_0 = \left( 2\pi \right)^{-i/2} e^{-(i+1)/12}
\label{eq:C0}
\end{equation}
(see, e.g., \cite[Appendix I]{bur_flipping}).

\section{The worst case error correction of regular GLDPC codes} \label{sec:regular_GLDPC}

In this paper we analyze the iterative parallel bit flipping bounded distance decoding algorithm \cite{tanner,chilappagari2010trapping}, applied to a random code in the regular GLDPC ensemble. The algorithm is an extension of the parallel bit flipping decoding algorithm for LDPC codes \cite{galmono,zyablov,sipser} to the GLDPC case.

Denote by $\bw = {\rm BDD}(\bv,t)$ the result of $t$-bounded distance decoding of the vector $\bv$ according to the code $\cC_0$ that has minimum distance at least $2t+1$ for $t>0$. If there exists a codeword $\bc\in\cC_0$, at Hamming distance at most $t$ from $\bv$, then $\bw=\bc$. Otherwise, $\bw = {\rm NULL}$.
Also denote by $\bv^{(l-1)}$ the $N$-dimensional vector of values of the variable nodes of the GLDPC code before the $l$-th iteration, $l=1,2,\ldots$. In particular, $\bv^{(0)}$, which is the input to the decoding algorithm, is used to initialize the variable nodes. 
Let $\cN(j,m)$, $1\le m\le d$, be the $m$-th variable node neighbor of check node $j$ in the Tanner graph. Similarly, $\cN(i,m)$, $1\le m\le c$, is the $m$-th check node neighbor of variable node $i$ in the Tanner graph.
We define $\bv_j^{(l-1)}$ to be the $d$ dimensional vector of values of variable nodes connected to check node $j$ before the $l$-th iteration, such that the $m$-th component, $v_{j,m}^{(l-1)}$, is the value of the $m$-th variable node neighbor of check node $j$, $\cN(j,m)$.
The parallel bit flipping bounded distance decoding algorithm is defined in Algorithm \ref{alg:BDD}.
Essentially, in the $l$ iteration step, each check node $j$ performs bounded distance decoding on $\bv_j^{(l-1)}$, obtained from the values transmitted to $j$ from its neighbors. If the bounded distance decoding obtains a codeword, $\bw_j^{(l-1)}\in \cC_0$, then check node $j$ sends a flip message to its variable node neighbors, $i$, for which there is a disagreement between their current value and the corresponding value of $\bw_j^{(l-1)}$. Afterwards, all variable nodes that obtained at least $c_1$ flip messages, flip their value.

There are some differences between the binary and non-binary cases. The differing parts that apply to the non-binary case are indicated in square brackets in Algorithm \ref{alg:BDD}.
The components of the vectors $\bv^{(l)}$, $\bv_j^{(l)}$ and $\bw_j^{(l)}$ take values in $\{0,1,\ldots,q-1\}$ ($\{0,1\}$ for the binary case where $q=2$). When the algorithm is used for decoding non-binary GLDPC codes ($q>2$), rather than a ``flip'' message we use a ``flip to $a$'' message where $a$ is the value to flip to.

\begin{algorithm}
\caption{Parallel bit flipping bounded distance decoding for GLDPC codes (the non-binary case is shown in square brackets)} \label{alg:BDD}
\begin{algorithmic}
\Require $\bv^{(0)}$, an $N$-dimensional vector, each component in $\{0,1\}$ [$\{0,1,\ldots,q-1\}$].
\State {\bf Initialization:} Set the initial values of the variables to $\bv^{(0)}$.
\For{$l=1,2,\ldots$} \Comment{In $l$-th iteration we determine $\bv^{(l)}$ from $\bv^{(l-1)}$}
	\State Each variable node sends its current value to all its check node neighbors.
	\For{each check node, $j\in\cJ$,} \Comment{Check nodes update}
		\State Compute $\bw_j^{(l-1)} = {\rm BDD}(\bv_j^{(l-1)},t)$.
		\If {$\bw_j^{(l-1)} \ne {\rm NULL}$}
			\For{$m=1,2,\ldots,d$,}
				\State $i=\cN(j,m)$.
				\State $a=w_{j,m}^{(l-1)}$.
				\If {$a\ne v_{i}^{(l-1)}$}
					\State Check node $j$ sends a flip [flip to $a$] message to variable node $i$.
				\EndIf
			\EndFor
		\EndIf
	\EndFor
	\For {each variable node, $i\in\cV$,} \Comment{Variable nodes update}
		\If {$i$ received at least $c_1$ flip [flip to $a$] messages}
			\State $v_i^{(l)} = \overline{v}_i^{(l-1)}$ [$v_i^{(l)} = a$] \Comment{flip variable}
		\Else
			\State $v_i^{(l)} = v_i^{(l-1)}$
		\EndIf
	\EndFor
\EndFor
\end{algorithmic}
\end{algorithm}

{\bf Notes:}
\begin{enumerate}
\item 
The vector $\bv^{(0)}$ can be obtained either directly from the respective channel output vector, or from some initial decoding scheme (e.g., belief propagation (BP) decoding).
\item
We iterate until the algorithm is stuck or until $l=T$, where $T$ is some pre-specified constant, whichever comes sooner.
\item
$c_1\in[1,c]$ is a parameter of the algorithm. In the non-binary case we impose the constraint $c_1>c/2$ since otherwise the same variable node might receive both $c_1$ ``flip to $a$'' and $c_1$ ``flip to $b$'' messages for $a\ne b$.
\item
When parallel edges connect some pair of variable and check nodes, we count a flip vote of the check node according to the multiplicity of the parallel edges (each edge has one vote).
\end{enumerate}

Following~\cite{sipser} we say that a variable node is {\em corrupt} if its value is different from the true transmitted codeword bit value. Otherwise, we say that the variable node is {\em correct}.
Following~\cite{sipser, chilappagari2010trapping} we say that a check node is {\em confused} if it sends flip messages to one or more correct variable nodes, or if it does not send flip messages to one or more corrupt variable nodes. Otherwise, the check node is said to be {\em helpful}.

\begin{figure}[htbp]
\centering
\includegraphics[width=0.6\linewidth]{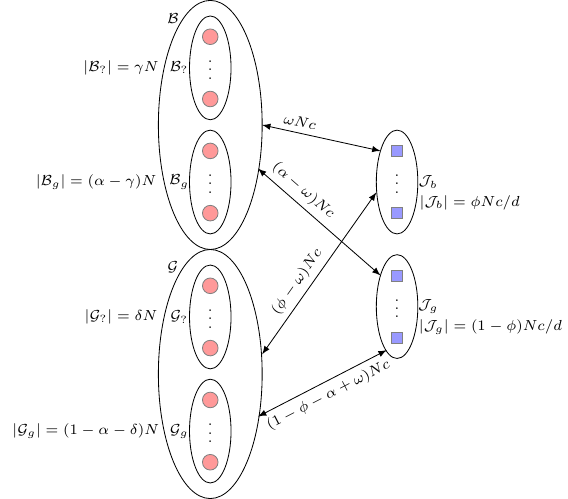} 
\caption{The relation between the various sets.
The sets of corrupt and correct variables before the iteration are denoted by $\cB$ and $\cG$.
Each element of $\cJ_b$ has more than $t$ edges connecting it to $\cB$.
Each element of $\cJ_g$ has up to $t$ edges connecting it to $\cB$.
Each element of $\cB_?$ ($\cB_g$, respectively) is connected to $\cJ_g$ by less than (at least) $c_1$ edges.
Each element of $\cG_?$ ($\cG_g$, respectively) is connected to $\cJ_b$ by at least (less than) $c_1$ edges.
After the iteration, each $v\in \cB_g$ will become correct, and each $v\in\cG_g$ will stay correct.
\label{fig:sets}}
\end{figure}

We now analyze the error correction capability of the algorithm.
In this section we consider binary GLDPC codes, and in Section \ref{sec:nonbinary} we extend to non-binary GLDPC codes. 
Consider some iteration of the algorithm. We partition the set $\cV$ of variable nodes into $\cV = \cB \cup \cG$, where $\cB$ and $\cG$ are the sets of corrupt and correct variable nodes at the beginning of the iteration.
We further partition $\cJ$ into $\cJ = \cJ_b \cup \cJ_g$, where $\cJ_b$ and $\cJ_g$ are defined as follows.
Each element of $\cJ_b$ has more than $t$ edges connecting it to $\cB$ (the other edges are connected to $\cG$).
Each element of $\cJ_g$ has up to $t$ edges connecting it to $\cB$ (the other edges are connected to $\cG$).
Hence, each element of $\cJ_g$ is helpful: It sends only correct decision messages (to flip or not to flip) to its variable node neighbors. That is, it sends flip (not to flip, respectively) message to each neighbor in $\cB$ ($\cG$). The elements of $\cJ_b$ are possibly confused.
We also partition $\cB$ into $\cB = \cB_? \cup \cB_g$ and $\cG$ into $\cG = \cG_? \cup \cG_g$, where $\cB_?$, $\cB_g$, $\cG_?$ and $\cG_g$ are defined as follows.
Each element of $\cB_?$ ($\cB_g$, respectively) is connected to $\cJ_g$ by less than (at least) $c_1$ edges.
Each element of $\cG_?$ ($\cG_g$, respectively) is connected to $\cJ_b$ by at least (less than) $c_1$ edges.
Thus, after the iteration, each $v\in \cB_g$ will become correct, and each $v\in\cG_g$ will stay correct.
Hence, the total number of corrupt variables after the iteration is at most $|\cB_?|+|\cG_?|$.
The relation between the various sets is depicted in Fig.~\ref{fig:sets}.

We denote the cardinalities of the various sets as follows:
\begin{equation}
\label{eq:card1}
|\cJ_b|   \defined \phi N c / d
\quad , \quad
|\cJ_g|   \defined (1-\phi) N c / d
\end{equation}
so that $J = |\cJ| = |\cJ_b| + |\cJ_g| = N c / d$. Also,
\begin{equation}
\label{eq:card2}
|\cB|   \defined \alpha N
\quad , \quad
|\cG|   \defined (1-\alpha) N
\end{equation}
\begin{equation}
\label{eq:card3}
|\cB_?| \defined \gamma N
\quad , \quad
|\cB_g| = (\alpha-\gamma) N
\quad , \quad
|\cG_?| \defined \delta N
\quad , \quad
|\cG_g| = (1-\alpha-\delta) N \: .
\end{equation}
We further denote by $\omega N c$ the number of edges connecting $\cB$ and $\cJ_b$.
Hence, there are $(\alpha-\omega)N c$ edges connecting $\cB$ and $\cJ_g$,
there are $(\phi-\omega) N c$ edges connecting $\cG$ and $\cJ_b$, and there are
$(1-\phi-\alpha+\omega) N c$ edges connecting $\cG$ and $\cJ_g$.

\begin{definition}
\label{def:badconfig}
Let $c_1$ and $t$ be some fixed positive integers.
Consider a $(c,d)$-regular bipartite graph with $N$ variable nodes, $\cV$, and $J=Nc/d$ check nodes, $\cJ$. Let $(\cB,\cG)$ be a partition of $\cV=\cB \cup \cG$, $(\cJ_b,\cJ_g)$ a partition of $\cJ = \cJ_b \cup \cJ_g$, $(\cB_?,\cB_g)$ a partition of $\cB = \cB_? \cup \cB_g$, and $(\cG_?,\cG_g)$ a partition of $\cG = \cG_? \cup \cG_g$.
We say that the partitioned bipartite graph,
$G(\cB, \cB_?, \cB_g, \cG, \cG_?, \cG_g, \cJ_b, \cJ_g)$,
is a $\Gamma(\alpha N, \gamma N, \delta N, \phi N c / d, \omega N c, c_1, t)$-partitioned bipartite graph, if it satisfies the following conditions:
\begin{enumerate}
	\item Each element of $\cJ_b$ has more than $t$ edges connecting it to $\cB$.
	\item Each element of $\cJ_g$ has up to $t$ edges connecting it to $\cB$.
	\item Each element of $\cB_?$ ($\cB_g$, respectively) is connected to $\cJ_g$ by less than (at least) $c_1$ edges.
	\item Each element of $\cG_?$ ($\cG_g$, respectively) is connected to $\cJ_b$ by at least (less than) $c_1$ edges.
	\item The cardinalities of the various sets are given by \eqref{eq:card1}--\eqref{eq:card3} for integer $\alpha N \ge 1$, $\gamma N$, $\delta N$ and $\phi N c / d$.
	\item There are $\omega N c$ edges connecting $\cB$ and $\cJ_b$.
\end{enumerate}
\end{definition}
Thus, $\Gamma(\alpha N, \gamma N, \delta N, \phi N c / d, \omega N c, c_1, t)$ is the set of all partitioned bipartite graphs that satisfy the relations depicted in Fig.~\ref{fig:sets}.
Recall that the total number of corrupt variables after the iteration is at most $|\cB_?|+|\cG_?|$, while there are $|\cB|=|\cB_?|+|\cB_g|$ corrupt variables before the iteration. Also recall \eqref{eq:card3}. Hence, we say that if
\begin{equation}
\label{eq:bad}
\delta \ge (\alpha-\gamma)
\end{equation}
then a $\Gamma(\alpha N, \gamma N, \delta N, \phi N c / d, \omega N c, c_1, t)$-partitioned bipartite graph,
$G(\cB, \cB_?, \cB_g, \cG, \cG_?, \cG_g, \cJ_b, \cJ_g)$, is {\em possibly bad}, since the number of errors after the iteration may be larger than or equal to the number of errors before the iteration.

By the definitions above, the following relations must hold (see Fig. \ref{fig:sets}),
\begin{equation}
\label{eq:gamma_bnd}
0 \le \gamma \le \alpha, \quad
0 \le \delta \le (1-\alpha), \quad
0 \le \omega c \le \min(\alpha,\phi) c, \quad
(\alpha-\omega) c \le (1-\phi) c \: .
\end{equation}
In addition, since each element of $\cJ_b$ has more than $t$ edges connecting it to $\cB$, it follows that $\omega N c \ge (t+1) \phi N c /d$. That is,
\begin{equation}
\label{eq:pi_bnd}
\phi \le \frac{\omega d}{t+1} \le \frac{\alpha d}{t+1} \: .
\end{equation}
Since each element of $G_?$ is connected to $\cJ_b$ by at least $c_1$ edges, it follows that
$\delta N c_1 \le (\phi-\omega) N c$. Hence,
\begin{equation}
\label{eq:delta_bnd}
\delta
\le \frac{(\phi-\omega)c}{c_1}
\le \frac{\phi c}{c_1}
\le \frac{\alpha d c}{(t+1)c_1}
\end{equation}
where the last inequality is due to \eqref{eq:pi_bnd}.
Also, since each element of $\cB_?$ is connected to $\cJ_b$ by at least $c-c_1+1$ edges, it follows that $\gamma N (c-c_1+1) \le \omega N c$. That is,
\begin{equation}
\label{eq:gamma_bnd1}
\gamma
\le \frac{\omega c}{c-c_1+1} \: .
\end{equation}
By~\eqref{eq:gamma_bnd}-\eqref{eq:delta_bnd},
\begin{equation}
\gamma = O(\alpha), \:
\phi = O(\alpha), \:
\omega = O(\alpha), \:
\delta = O(\alpha)
\label{eq:o_alpha}
\end{equation}
where $a = O(\alpha)$ means that there exists some constant $C$ such that $a < C \alpha$.

To formulate our first theorem we define the following polynomials,
\begin{equation}
F_{0}(x) \defined \sum_{j=c-c_1+1}^{c} \binom{c}{j} x^j,
\quad F_{1}(x) \defined \sum_{j=0}^{c-c_1} \binom{c}{j} x^j,
\quad F_{2}(x) \defined \sum_{j=c_1}^{c} \binom{c}{j} x^j,
\quad F_{3}(x) \defined \sum_{j=0}^{c_1-1} \binom{c}{j} x^j \label{eq:F0} \\
\end{equation}
\begin{equation}
G_0(x) \defined \sum_{j=t+1}^d \binom{d}{j} x^j,
\quad G_1(x) \defined \sum_{j=0}^t \binom{d}{j} x^j \: .
\label{eq:G1}
\end{equation}
These polynomials will be used to formulate various generating functions.
We also define,
\begin{align}
\label{eq:t1}
t_1 &\defined \inf_{x>0} \left\{ \gamma \log F_0(x) + (\alpha - \gamma) \log F_1(x) -
\omega c \log x \right\}
\\
\label{eq:t2}
t_2 &\defined \inf_{x>0} \left\{ \delta \log F_2(x) + (1 - \alpha - \delta) \log F_3(x) -
(\phi-\omega)c \log x \right\}
\\
\label{eq:u1}
u_1 &\defined \inf_{x>0} \left\{ \frac{\phi c}{d} \log G_0(x) - \omega c \log x \right\}
\\
\label{eq:u2}
u_2 &\defined \inf_{x>0} \left\{ \frac{(1-\phi) c}{d} \log G_1(x) - (\alpha-\omega) c \log x \right\}
\\
\label{eq:rho}
\rho &\defined t_1 + t_2 + u_1 + u_2
\end{align}
\begin{equation}
\psi(\alpha,\gamma,\delta,\phi,\omega)
\defined
h(\gamma,\alpha-\gamma,\delta)
+ (c/d) h(\phi)
+
\rho - c h(\omega,\alpha-\omega,\phi-\omega)
\label{eq:psi}
\end{equation}
where $h()$ is the entropy function defined in~\eqref{eq:entropy}, and
\begin{equation}
\label{eq:falpha}
f(\alpha) \defined \max_{\gamma,\delta,\phi,\omega} \psi(\alpha,\gamma,\delta,\phi,\omega)
\: .
\end{equation}
The last maximization is over values of $\gamma,\delta,\phi$ and $\omega$ that satisfy~\eqref{eq:bad}-\eqref{eq:gamma_bnd1}.

Our main result in this section is the following:
\begin{theorem}
\label{thrm:regular_GLDPC}
Consider the ensemble of $(c, d)$-regular binary GLDPC codes with blocklength $N$. Each component code has blocklength $d$ and it can correct all error patterns of size $t>0$ or less.
Let $\alpha_0>0$ be the smallest positive root (we show that such root does exist) of the function $f(\alpha)$ defined in \eqref{eq:falpha}.
Suppose that the parallel bit flipping bounded distance decoding algorithm is used to decode codes from this ensemble under the conditions,
\begin{equation}
c_1 \ge 2, 
\quad
(c-c_1+1)\cdot \frac{t}{t+1} > 1 \: .
\label{eq:c1_cond}
\end{equation}
Then, for any $\overline{\alpha}_0<\alpha_0$ and $N$ sufficiently large, almost all codes in this ensemble (i.e., except for a negligible fraction of codes) satisfy the following:
Any transmitted codeword corrupted by any error pattern of size at most $\lfloor \overline{\alpha}_0 N \rfloor$ will be decoded correctly.

Note that the condition \eqref{eq:c1_cond} holds for all $c\ge 4$ and $t\ge 1$, and also for $c=3$ and all $t\ge 2$ (e.g., select $c_1=\lceil c/2 \rceil$).
\end{theorem}
The theorem asserts, for $N$ sufficiently large, under the condition \eqref{eq:c1_cond}, that a simple random selection of a Tanner graph yields, with probability approaching one, a good code that can correct a constant fraction of worst case errors.

\beginproof
To prove the theorem we consider a random code (graph) drawn from the ensemble. Using the union bound we show, for $N$ sufficiently large, that with high probability there are no possible $\Gamma(\alpha N, \gamma N, \delta N, \phi N c / d, \omega N c, c_1, t)$-partitioned bipartite graphs that satisfy $\alpha < \alpha_0$ and \eqref{eq:bad}-\eqref{eq:gamma_bnd1}.
Hence, for almost all codes in the ensemble, there are no possibly bad $\Gamma(\alpha N, \gamma N, \delta N, \phi N c / d, \omega N c, c_1, t)$-partitioned bipartite graphs with $\alpha < \alpha_0$, so that if we start with arbitrary $\alpha N\ge 1$ errors where $\alpha < \alpha_0$, each iteration strictly decreases the number of errors until all errors have been corrected.

Denote by $\overline{p}_e(\lfloor \overline{\alpha}_0 N \rfloor)$, the probability over our ensemble that there exists a possibly bad $\Gamma(\alpha N, \gamma N, \delta N, \phi N c / d, \omega N c, c_1, t)$-partitioned bipartite graph with $|\cB| = \alpha N$ for some positive integer $\alpha N$ that satisfies $\alpha N \le \lfloor \overline{\alpha}_0 N \rfloor$.
That is, $\overline{p}_e(\lfloor \overline{\alpha}_0 N \rfloor)$ is the fraction of bad codes in the $(c, d)$-regular GLDPC ensemble that cannot correct all combinations of at most $\lfloor \overline{\alpha}_0 N \rfloor$ errors. We will show that $\overline{p}_e(\lfloor \overline{\alpha}_0 N \rfloor) \rightarrow 0$ as $N\rightarrow\infty$, thus proving the theorem.

By the union bound,
\begin{align}
\nonumber
\overline{p}_e(\lfloor \overline{\alpha}_0 N \rfloor)
&=
\Pr \left\{ \exists \: \left( \cB, \cB_?, \cB_g, \cG, \cG_?, \cG_g, \cJ_b, \cJ_g \right) \: : \right.\\
&\quad \quad \: \: \: \left. G(\cB, \cB_?, \cB_g, \cG, \cG_?, \cG_g, \cJ_b, \cJ_g) \in
\Gamma(\alpha N, \gamma N, \delta N, \phi N c / d, \omega N c, c_1, t)
\right.
\nonumber \\
&\quad \quad \: \: \left. \mbox{ where $\alpha N \le \lfloor \overline{\alpha}_0 N \rfloor$, such that~\eqref{eq:bad} holds}
\right\}
\nonumber \\
&\le
\sum_{\alpha N \le \lfloor \overline{\alpha}_0 N \rfloor} p_e(\alpha N)
\label{eq:pe}
\end{align}
where
\begin{equation}
\label{eq:pe_alpha}
p_e(\alpha N) \defined
\sum_{\gamma N, \delta N, \phi N c/d, \omega N c}
\frac{\eta \cdot \zeta}{(Nc)!} \: .
\end{equation}
The summation in~\eqref{eq:pe} is over positive integer values of $\alpha N$ such that $\alpha N \le \lfloor \overline{\alpha}_0 N \rfloor$. The summation in~\eqref{eq:pe_alpha} is over non-negative integer values of $\gamma N, \delta N, \phi N c/d$ and $\omega N c$ that satisfy~\eqref{eq:bad}-\eqref{eq:gamma_bnd1} (since~\eqref{eq:bad} holds, we enumerate only possibly bad $\Gamma(\alpha N, \gamma N, \delta N, \phi N c / d, \omega N c, c_1, t)$-partitioned bipartite graphs).

The term $\eta$ is the number of ways to partition the set of variable nodes into sets of cardinalities $\gamma N$, $(\alpha-\gamma)N$, $\delta N$ and $(1-\alpha-\delta)N$, and to partition the set of check nodes into sets of cardinalities $\phi N c/d$ and $(1-\phi)N c/d$. It is given by
(note our notation~\eqref{eq:choose}),
\begin{equation}
\label{eq:eta}
\eta =
\binom{N}{\gamma N, (\alpha-\gamma)N, \delta N}
\cdot \binom{N c/d}{\phi N c/d} \: .
\end{equation}
The term $\zeta$ is the number of possible graphs in which, for a given fixed partition, $\cB, \cB_?, \cB_g, \cG, \cG_?, \cG_g, \cJ_b, \cJ_g$,
\begin{equation}
\label{eq:G_in_Gamma}
G(\cB, \cB_?, \cB_g, \cG, \cG_?, \cG_g, \cJ_b, \cJ_g)
\in
\Gamma(\alpha N, \gamma N, \delta N, \phi N c / d, \omega N c, c_1, t) \: .
\end{equation}
The total number of unrestricted graphs in the ensemble is $(Nc)!$ (number of all possible ways to match left and right sockets). Thus, $\zeta / (Nc)!$ is the probability that, for a given fixed partition, $\cB, \cB_?, \cB_g, \cG, \cG_?, \cG_g, \cJ_b, \cJ_g$, \eqref{eq:G_in_Gamma} holds.

Now, $\eta$ can be bounded by using~\eqref{eq:multi_bnd_1}, yielding
\begin{equation}
\label{eq:eta_bnd_1}
\eta \le \exp \left\{ N \left[
h(\gamma,\alpha-\gamma,\delta) + (c/d) h(\phi)
\right] \right\}
\end{equation}
where the entropy function $h()$ is defined in~\eqref{eq:entropy}.

We now turn to the other term, $\zeta$. First note that
\begin{equation}
\label{eq:zeta}
\zeta =
\zeta_l \cdot \zeta_r \cdot (\omega N c)! ((\alpha-\omega) N c)! [(\phi-\omega)N c]! [(1-\phi-\alpha+\omega) N c]!
\end{equation}
where $\zeta_l$ is the number of ways to choose $\omega N c$ and $(\phi-\omega) N c$ left sockets in $\cB$ and $\cG$ respectively, that connect to $\cJ_b$ (i.e., number of ways to mark left sockets by either $\cJ_b$ or $\cJ_g$).
Similarly, $\zeta_r$ is the number of ways to choose $\omega N c$ and $(\alpha-\omega) N c$ right sockets in $\cJ_b$ and $\cJ_g$, respectively, that connect to $\cB$ (i.e., number of ways to mark right sockets by either $\cB$ or $\cG$).

Now, $\zeta_l$ and $\zeta_r$ can be expressed using generating functions as follows.
Denote by $[x^k]F(x)$ the coefficient of $x^k$ in the polynomial $F(x)$. Then (note, by the definition of $\cB_?$ and $\cB_g$, that each $v\in\cB_?$ ($\cB_g$, respectively) is connected to $\cJ_b$ by at least (less than) $c-c_1+1$ edges),
\begin{equation}
\label{eq:zeta_l}
\zeta_l =
\left[ x^{\omega N c} \right]
\left(
F_0(x)^{\gamma N} F_1(x)^{(\alpha - \gamma)N}
\right)
\cdot
\left[ x^{(\phi-\omega) N c} \right]
\left(
F_2(x)^{\delta N} F_3(x)^{(1 - \alpha - \delta)N}
\right)
\end{equation}
and
\begin{equation}
\label{eq:zeta_r}
\zeta_r
=
\left[ x^{\omega N c} \right] G_0(x)^{\phi N c/d}
\cdot
\left[ x^{(\alpha-\omega) N c} \right] G_1(x)^{(1-\phi) N c/d}
\end{equation}
where the polynomials $F_0(x), F_1(x), F_2(x), F_3(x), G_0(x)$ and $G_1(x)$ are defined in~\eqref{eq:F0}-\eqref{eq:G1}.

The saddle point inequality (see e.g.~\cite{asympt_enum} and references therein) asserts that, for a polynomial $F(x)$ with non-negative coefficients,
\begin{equation}
[x^k]F(x) \le 
\inf_{x>0} \frac{F(x)}{x^k} =
\exp \left\{
\inf_{x>0} \left[ \log F(x) - k\log x \right]
\right\} \: .
\label{eq:saddle_point_ineq}
\end{equation}
Applying \eqref{eq:saddle_point_ineq} to \eqref{eq:zeta_l} yields,
\begin{equation}
\label{eq:zeta_l_bnd}
\zeta_l
\le
\exp \left\{ N \left[ t_1 + t_2 \right]\right\}
\end{equation}
where $t_1$ and $t_2$ are defined in~\eqref{eq:t1} and~\eqref{eq:t2}.
Similarly, applying \eqref{eq:saddle_point_ineq} to \eqref{eq:zeta_r} yields,
\begin{equation}
\label{eq:zeta_r_bnd}
\zeta_r
\le
\exp \left\{ N \left[ u_1 + u_2 \right]\right\}
\end{equation}
where $u_1$ and $u_2$ are defined in~\eqref{eq:u1} and~\eqref{eq:u2}.
By~\eqref{eq:zeta},~\eqref{eq:zeta_l_bnd},~\eqref{eq:zeta_r_bnd}, \eqref{eq:rho} and~\eqref{eq:multi_upbnd} we obtain,
\begin{align}
\nonumber
\frac{\zeta}{[Nc]!}
&\le
e^{\rho N} \cdot
\binom{cN}{\omega c N, (\alpha-\omega) c N, (\phi-\omega)c N}^{-1}
\\
&\le
C_1 \cdot \left( N c \right)^{3/2} \sqrt{\omega (\alpha-\omega)(\phi-\omega)} \cdot
\exp \left\{ N \left[ \rho - c h(\omega,\alpha-\omega,\phi-\omega)
\right] \right\}
\label{eq:zeta_bnd1a}
\end{align}
for $C_1=1/C_0$ where $C_0$ is given in \eqref{eq:C0}.
Combining this inequality with \eqref{eq:eta_bnd_1} we obtain the following upper bound on the summation term in \eqref{eq:pe_alpha},
\begin{equation}
\label{eq:eta_zeta_bnd}
\frac{\eta \cdot \zeta}{[Nc]!}
\le
C_1 \cdot \left( N c \right)^{3/2} \sqrt{\omega (\alpha-\omega)(\phi-\omega)} \cdot
\exp \left\{
N \left[ 
\rho - c h(\omega,\alpha-\omega,\phi-\omega) + h(\gamma,\alpha-\gamma,\delta) + (c/d) h(\phi)
\right] \right\} \: .
\end{equation}

Now, by \eqref{eq:pe_alpha} and \eqref{eq:gamma_bnd}-\eqref{eq:gamma_bnd1},
\begin{align}
p_e(\alpha N)
&\le
(\alpha N + 1) \cdot \left( \frac{d c}{(t+1)c_1} \cdot \alpha N + 1 \right)
\cdot \left( \frac{c}{t+1} \cdot \alpha N  + 1 \right) \cdot \left(c \cdot \alpha N + 1 \right)
\cdot
\max_{\gamma N, \delta N, \phi N c / d, \omega N c}
\left\{
\frac{\eta \cdot \zeta}{(Nc)!}
\right\}
\nonumber\\ &=
C_2 \cdot \left( \alpha N \right)^4
\cdot
\max_{\gamma N, \delta N, \phi N c / d, \omega N c}
\left\{
\frac{\eta \cdot \zeta}{(Nc)!}
\right\}
\label{eq:pe_alpha_bnd}
\end{align}
where $C_2$ is a constant (it depends only on $c$, $c_1$, $d$ and $t$),
and where the maximization is over $\gamma N, \delta N, \phi N c / d$ and $\omega N c$ that satisfy \eqref{eq:bad}-\eqref{eq:gamma_bnd1}.
Incorporating \eqref{eq:eta_zeta_bnd} in \eqref{eq:pe_alpha_bnd} yields (noting that, by \eqref{eq:gamma_bnd}-\eqref{eq:pi_bnd}, $\omega\le\alpha$, $\alpha-\omega\le\alpha$ and $\phi-\omega \le \phi \le \alpha d / (t+1)$),
\begin{equation}
p_e(\alpha N)
\le
C_3 \cdot \left( \alpha N \right)^{11/2}
\cdot
\exp\left\{
N \cdot \max_{\gamma, \delta, \phi, \omega} \psi(\alpha,\gamma,\delta,\phi,\omega)
\right\}
\label{eq:pe_alpha_bnd1}
\end{equation}
where $C_3$ is a constant (it depends only on $c$, $c_1$, $d$ and $t$) and $\psi()$ is defined in \eqref{eq:psi}.
The maximization is over values of $\gamma$, $\delta$, $\phi$ and $\omega$ that satisfy~\eqref{eq:bad}-\eqref{eq:gamma_bnd1}.
By the definition of $f(\alpha)$ in~\eqref{eq:falpha}, we can rewrite~\eqref{eq:pe_alpha_bnd1} as,
\begin{equation}
\label{eq:pe_alpha_bndf}
p_e(\alpha N)
\le
C_3 \cdot \left( \alpha N \right)^{11/2} e^{N f(\alpha)} \: .
\end{equation}

Next, we consider the upper bound \eqref{eq:pe_alpha_bndf} for small values of $\alpha$. First note that for $a\in(0,1)$ that satisfies $a=O(\alpha)$ (i.e., $a<C\alpha$ for some constant $C$), we have $a \log (a/\alpha) = O(\alpha)$. Hence, $a\log a = a \log \alpha + O(\alpha)$. In addition, for such $a$, we also have $-(1-a)\log(1-a) = O(\alpha)$ (due to the inequality $\log (1-a)\ge -a/(1-a)$, $\forall a<1$). Hence, by the definition of $\psi()$ in \eqref{eq:psi} and by \eqref{eq:o_alpha},
\begin{equation}
\psi(\alpha,\gamma,\delta,\phi,\omega)
\le
\left[ -(\gamma+\alpha-\gamma+\delta) - (c/d)\phi + c(\omega+\alpha-\omega+\phi-\omega) \right]
\log \alpha
+
\rho + O(\alpha)
\label{eq:hpsi_ineq}
\end{equation}
where $\rho$ is defined in \eqref{eq:rho}.
To bound $t_1$, we set $x=1$ in~\eqref{eq:t1}.
To bound $t_2$, we set $x=\alpha$ in~\eqref{eq:t2}.
To bound $u_1$, we set $x=1$ in~\eqref{eq:u1}.
To bound $u_2$, we set $x=\alpha$ in~\eqref{eq:u2}.
Using~\eqref{eq:o_alpha} we thus obtain,
\begin{align}
\label{eq:t1_bnd}
t_1 &\le O(\alpha)
\\
t_2 &\le \delta \log \left(\sum_{j=c_1}^{c} \binom{c}{j} \alpha^j \right) +
(1-\alpha-\delta) \log \left(\sum_{j=0}^{c_1-1} \binom{c}{j} \alpha^j \right)
- (\phi-\omega) c \log \alpha
\nonumber\\ &= \delta \log \left( \binom{c}{c_1} \alpha^{c_1} \right) +
(1-\alpha-\delta) \log \left( 1 + c\alpha \right)
- (\phi-\omega) c \log \alpha + O(\alpha)
\nonumber\\ &= \left[ \delta c_1 - \phi c + \omega c \right] \log \alpha + O(\alpha)
\label{eq:t2_bnd}
\\
\label{eq:u1_bnd}
u_1 &\le O(\alpha)
\\
u_2 &\le \frac{(1-\phi)c}{d} \log\left( 1 + d\alpha \right) - (\alpha-\omega) c \log \alpha + O(\alpha)
\nonumber\\ &= -(\alpha-\omega)c\log \alpha + O(\alpha) \: .
\label{eq:u2_bnd}
\end{align}

Combining \eqref{eq:rho} and \eqref{eq:hpsi_ineq}-\eqref{eq:u2_bnd} and collecting terms we conclude that,
\begin{align}
\psi(\alpha,\gamma,\delta,\phi,\omega)
&\le
\left[ -\alpha - \delta - (c/d)\phi + \alpha c + \phi c - \omega c + \delta c_1 - \phi c +
\omega c - \alpha c + \omega c \right] \log \alpha + O(\alpha)
\nonumber \\ &=
\left[ -\alpha + \delta (c_1 - 1) - (c/d)\phi + \omega c \right] \log \alpha + O(\alpha)
\nonumber \\
&\le
\left[
-\alpha + \delta \left( c_1 - 1 \right) + \omega c t / (t+1)
\right] \log \alpha + O(\alpha) \: .
\label{eq:psi_ineq_1}
\end{align}
In the last inequality we have used the fact that $(c/d) \phi \le \omega c / (t+1)$ which is due to \eqref{eq:pi_bnd} (recall that $\log \alpha < 0$).

Now, by \eqref{eq:pi_bnd} and \eqref{eq:delta_bnd},
\begin{equation}
\delta \le \frac{(\phi-\omega)c}{c_1} \le \frac{\omega c}{c_1} \left( \frac{d}{t+1} - 1 \right) \: .
\end{equation}
This relation can also be written as,
\begin{equation}
\label{eq:delta_bnd1}
\frac{\omega c t}{t+1} \ge \frac{t}{t+1} \cdot \frac{c_1 \delta}{d/(t+1)-1} \: .
\end{equation}
Next, we rewrite \eqref{eq:gamma_bnd1} as
\begin{equation}
\frac{\omega c t}{t+1} \ge
\frac{t}{t+1} \cdot (c-c_1+1) \gamma
\label{eq:gamma_bnd2}
\end{equation}
From \eqref{eq:delta_bnd1} and \eqref{eq:gamma_bnd2} we conclude that for any $\beta\in[0,1)$, we have,
\begin{equation}
\frac{\omega c t}{t+1} \ge \frac{t}{t+1} 
\left[ \beta \frac{c_1 \delta}{d/(t+1)-1} + (1-\beta) (c-c_1+1) \gamma \right] 
=
a_1(\beta) \delta + a_2(\beta) \gamma
\label{eq:omega_c_t_bnd}
\end{equation}
where
\begin{equation}
a_1(\beta) \defined \frac{c_1 t \beta}{d-t-1}
, \quad
a_2(\beta) \defined (c-c_1+1) (1-\beta) \cdot \frac{t}{t+1} \: .
\end{equation}

By \eqref{eq:psi_ineq_1}, \eqref{eq:omega_c_t_bnd} and \eqref{eq:bad},
\begin{align}
\psi(\alpha,\gamma,\delta,\phi,\omega)
&\le
\left[
-\alpha + \left( c_1 - 1 + a_1(\beta) \right) \delta + a_2(\beta) \gamma
\right] \log \alpha + O(\alpha)
\nonumber\\
&\le
\left[
-\alpha + (\delta+\gamma) \cdot \min \left[ c_1-1+a_1(\beta), a_2(\beta) \right]
\right] \log \alpha + O(\alpha)
\nonumber\\
&\le
\left[
-\alpha + \alpha\cdot\min \left[ c_1-1+a_1(\beta), a_2(\beta) \right] 
\right] \log \alpha + O(\alpha)
\nonumber\\
&=
\kappa \alpha \log \alpha + O(\alpha)
\end{align}
(we have used \eqref{eq:bad} in the third inequality). $\kappa$ is defined by,
\begin{equation}
\kappa \defined \min \left[ c_1-1+a_1(\beta), a_2(\beta) \right] - 1 \: .
\end{equation}
Recalling \eqref{eq:c1_cond} and \eqref{eq:d_t_ineq}, for $\beta>0$ sufficiently small,
\begin{equation}
c_1 - 1 + a_1(\beta) > 1, \quad a_2(\beta) > 1 \: .
\end{equation}
Hence $\kappa>0$.
Now, by the definition of $f(\alpha)$, \eqref{eq:falpha},
\begin{equation}
\label{eq:falpha_bnd}
f(\alpha) \le \kappa\alpha\log\alpha + O(\alpha) \: .
\end{equation}
Note that if $\alpha$ is sufficiently small, i.e., $\alpha\le\tilde{\alpha}$ for $\tilde{\alpha}$ small enough, then, since $\kappa>0$, the term $\kappa\alpha\log\alpha$ is negative and dominates (in absolute value) the other term, $O(\alpha)$. Thus
\begin{equation}
\label{eq:noa_neglect}
f(\alpha) \le
\kappa\alpha\log\alpha + O(\alpha) \le
\frac{1}{2} \kappa\alpha\log\alpha \: .
\end{equation}
Combining this result with \eqref{eq:pe_alpha_bndf} yields,
\begin{equation}
\label{eq:pe_alpha_bnd3}
p_e(\alpha N)
\le
C_3 \cdot \left( \alpha N \right)^{11/2} 
\exp \left\{ \frac{\alpha N \kappa}{2}\log\alpha \right\}
=
C_3 \cdot \left( \alpha N \right)^{11/2} \left( \frac{\alpha N}{N} \right)^{\alpha N \kappa/2}
\end{equation}
for $\alpha N = 1,2,\ldots,\tilde{\alpha} N$ where $\tilde{\alpha}$ is sufficiently small.
Denoting $i=\alpha N$ we have,
\begin{equation}
\label{eq:pe_i}
p_e(i)
\le
C_3 \cdot i^{11/2} \left( \frac{i}{N} \right)^{i \kappa/2}
\end{equation}
for $i=1,2,\ldots,\tilde{\alpha} N$.
Now, let $I$ be some fixed positive integer.
Then by~\eqref{eq:pe_i} (note that $(i/N)^{i\kappa/2} = {\rm exp}\{\kappa N / 2 \cdot\alpha\log \alpha\}$ and $\alpha\log\alpha$ is monotonically decreasing for $\alpha\in(0,1/e)$. Hence, $(i/N)^{i\kappa/2}$ is monotonically decreasing for $1\le i\le \tilde{\alpha} N < (1/e)N$),
\begin{equation}
\sum_{\alpha N = 1}^{\talpha N} p_e(\alpha N) \le
C_3 \cdot \sum_{i=1}^{I-1} i^{11/2} \left( \frac{i}{N} \right)^{i\kappa/2}
+ \:
C_3 \cdot \talpha N \cdot \left( \talpha N \right)^{11/2} \left( \frac{I}{N} \right)^{I\kappa/2} \: .
\label{eq:sum_pe}
\end{equation}
By selecting $I$ sufficiently large, we see that the right hand side of the last inequality approaches $0$ as $N\rightarrow\infty$. Also, by~\eqref{eq:pe} and~\eqref{eq:pe_alpha_bndf},
\begin{equation}
\overline{p}_e(\lfloor \overline{\alpha}_0 N \rfloor)
=
\sum_{\alpha N=1}^{\talpha N} p_e(\alpha N) +
\sum_{\alpha N=\talpha N + 1}^{\lfloor \overline{\alpha}_0 N \rfloor} p_e(\alpha N)
\le
\sum_{\alpha N=1}^{\talpha N} p_e(\alpha N) +
\sum_{\alpha N=\talpha N + 1}^{\lfloor \overline{\alpha}_0 N \rfloor} C_3 \cdot \left( \alpha N \right)^{11/2} e^{N f(\alpha)} \: .
\end{equation}
Now, both sums on the right hand side approach $0$ as $N\rightarrow\infty$: For the first sum this follows from \eqref{eq:sum_pe} as we argued above. For the second sum this is due to the fact that each term is dominated by a negative exponential in $N$. Hence,
$\overline{p}_e(\lfloor \overline{\alpha}_0 N \rfloor) \rightarrow 0$ as $N\rightarrow\infty$ as claimed.

By \eqref{eq:noa_neglect} we conclude that for a sufficiently small $\alpha>0$, $f(\alpha)<0$.
The number $\alpha_0$, which we defined as the smallest positive root of $f(\alpha)$, must obviously exist since otherwise, by the result that we have just proved, we would be able to correct arbitrary error patterns with an arbitrary number of errors.
\finproof
Notes:
\begin{enumerate}
\item 
The introduction of $\beta\in[0,1)$ in the proof is only required for the case where $c_1=2$. In all other cases we can simply set $\beta=0$. 
\item 
Since $f(\alpha)$ is defined as the maximum of a continuous function ($t_1$, $t_2$, $u_1$ and $u_2$ are continuous in $\alpha$, $\gamma$, $\delta$, $\omega$ and $\phi$, see \cite[Section III]{asympt_enum}) over a polyhedral domain, then $f(\alpha)$ is a continuous function.
\item Similarly to \cite[Remark 13]{sipser} and \cite{bur_flipping} (the comment after the proof of Theorem 1), we can simulate sequentially the parallel bit flipping bounded distance decoding algorithm in Theorem \ref{thrm:regular_GLDPC}, such that it achieves successful decoding in linear time.
\end{enumerate}

We conclude that for $N$ sufficiently large, we can correct a constant fraction of worst case errors if either
$c\ge 4$ and $t\ge 1$ or $c=3$ and $t\ge 2$.
We now compare this result to the one in \cite{chilappagari2010trapping}.
A $(c,d)$-regular bipartite Tanner graph is said to be a $(c,d,\alpha,\delta)$ expander, for some $\alpha>0$ and $\delta>0$, if for all sets $\cB$ of variable nodes with $|\cB|\le \alpha N$, the set of neighbors of $\cB$, denoted by $\cN(\cB)$, satisfies $|\cN(\cB)|>\delta c|\cB|$.
\cite[Theorem 4]{chilappagari2010trapping} considers a $(c,d)$-regular GLDPC code with blocklength $N$, where the component code can correct all error patterns of size $t>0$ or less. Suppose that the corresponding $(c,d)$-regular Tanner graph is a $(c,d,\alpha,\delta)$ expander with
\begin{equation}
\frac{t+2}{2(t+1)} < \delta < 1
\: .
\label{eq:t_delta}
\end{equation}
Then it is shown in \cite[Theorem 4]{chilappagari2010trapping} that the parallel bit flipping bounded distance decoding algorithm can correct all error patterns of size $\alpha N$ or less in any transmitted codeword.
However, the above theorem does not specify whether a graph with these expansion properties exists (and if it does exist, how to construct it).
Provably good expander graphs can be obtained with high probability by sampling at random from the $(c,d)$-regular bipartite Tanner graph ensemble, when the blocklength $N$ is sufficiently large \cite{sipser}.
Fix some values of $p\in(0,1)$ and $\delta<1-1/c$. The results on the expansion of a random graph drawn from the $(c,d)$-regular ensemble with blocklength $N$ \cite{bur_expander}[Lemma 1], \cite[Theorem 8.7]{ru_book}, \cite[Lemma 4]{bur_flipping} can be used to derive a sufficient lower bound on $N=N(p,\delta)$ such that the graph is a $(c,d,\alpha>0,\delta)$ expander with probability at least $p$. However, this sufficient lower bound on $N$ approaches infinity as $\delta$ approaches $1-1/c$ (from below) for any fixed $p\in (0,1)$.
In general, we cannot assert the existence of a $(c,d,\alpha>0,\delta)$ expander with $\delta\ge 1-1/c$.
In fact, by~\cite{ru_book}[Section 8.4], a $(c,d)$-regular graph cannot be a $(c,d,\alpha,\delta)$ expander, for $\alpha>0$ and $\delta>1-1/c$.
Combining this with \eqref{eq:t_delta} yields the requirement
\begin{equation}
\frac{t+2}{2(t+1)} < 1 - \frac{1}{c}
\: .
\end{equation}
Hence,
\begin{equation}
t > \frac{2}{c-2}
\: .
\label{eq:t_requirement}
\end{equation}
For $c=3$, \eqref{eq:t_requirement} asserts $t\ge 3$, and for $c=4$, \eqref{eq:t_requirement} asserts $t\ge 2$. Hence, Theorem \ref{thrm:regular_GLDPC} improves \cite[Theorem 4]{chilappagari2010trapping} for $c=3$ (since it only requires $t\ge 2$) and for $c=4$ (since it only requires $t\ge 1$).

The recent work \cite{cheng2024can} showed that $(2t+1)\delta>3$ is a sufficient condition for correcting a constant fraction of worst case errors in linear time. Now, for $c=3$, a graph chosen at random from the $(c,d)$-regular ensemble will be a $(c,d,\alpha,\delta)$ expander for some $\alpha>0$ and any $\delta<1-1/c=2/3$.
We can thus conclude that for $c=3$ it is sufficient to require $2t+1>9/2$, i.e., $t\ge 2$, in order to decode a constant fraction of worst case errors in linear time (we cannot conclude from \cite{cheng2024can} that $t\ge 1$ is sufficient for correcting a constant fraction of worst case errors in linear time when $c=4$). However, the result in \cite{cheng2024can} requires a much more complicated decoding algorithm with a much higher decoding time. In fact, the decoder in \cite{cheng2024can} activates the parallel bit flipping bounded distance decoding algorithm for a large (but constant) number of times, under an appropriate control logic. On the other hand, our result only uses a single run of the simple parallel bit flipping bounded distance decoding algorithm.

\section{Extensions} \label{sec:extensions}
\subsection{Random error correction} \label{sec:random_errors}
We have analyzed the ability of GLDPC codes to correct all error patterns with Hamming weight $\alpha N$ or less. Now suppose we wish to correct $\alpha N$ errors whose bit locations are determined uniformly at random (as in the case where we transmit over a discrete memoryless channel). In this case the error correction radius will be larger. To obtain an improved (compared to the previous case of worst case errors) bound, the first iteration is analyzed differently than the following ones. Denote by $p_e(\alpha N, \nu N)$ the probability over our ensemble that if we start the first iteration with a set of $\alpha N$ corrupt variables, chosen uniformly at random, then after the iteration there will be at least $\nu N$ corrupt variables.
Due to the symmetry with respect to variables in the definition of the ensemble, when calculating $p_e(\alpha N, \nu N)$, we can assume, without loss of generality, that the corrupt variables, $\cB$, are the first $\alpha N$ ones.
Also, recall that the total number of corrupt variables after the iteration is at most $|\cB_?|+|\cG_?| = (\gamma + \delta) N$. Hence,
\begin{align}
\nonumber
p_e(\alpha N, \nu N) &\le \Pr \left\{ \exists \: \cB_?, \cB_g, \cG_?, \cG_g, \cJ_b, \cJ_g \: : \right.\\
&\quad \quad \: \: \: \left.
G( \cB, \cB_?, \cB_g, \cG, \cG_?, \cG_g, \cJ_b, \cJ_g) \in
\Gamma(\alpha N, \gamma N, \delta N, \phi N c / d, \omega N c, c_1, t)
\right.
\\
&\quad \quad \: \: \: \left. \mbox{where $\gamma N + \delta N \ge \nu N$}
\right\}
\nonumber \\
&\le
\sum_{\gamma N, \delta N, \phi N c/d, \omega N c}
\frac{\tilde{\eta} \cdot \zeta}{(Nc)!} \: .
\label{eq:pe_r}
\end{align}
The second inequality is due to the union bound.
The summation in~\eqref{eq:pe_r} is over non-negative integer values of $\gamma N, \delta N, \phi N c/d$ and $\omega N c$ that satisfy~\eqref{eq:gamma_bnd}-\eqref{eq:gamma_bnd1} and $\gamma+\delta \ge \nu$.

The term $\tilde{\eta}$ is the number of ways to partition the set of corrupt variables (first $\alpha N$ variables) into two sets of $\gamma N$ and $(\alpha-\gamma)N$, to partition the set of correct variable nodes (last $(1-\alpha) N$ variables) into two sets of $\delta N$ and $(1-\alpha-\delta)N$, and to partition the set of check nodes into two sets of $\phi N c/d$ and $(1-\phi)N c/d$. It is given by,
\begin{equation}
\label{eq:eta_r}
\tilde{\eta} =
\binom{\alpha N}{\gamma N} \cdot
\binom{(1-\alpha)N}{\delta N} \cdot
\binom{N c/d}{\phi N c/d} \: .
\end{equation}
The term $\zeta$ is given by the same expression \eqref{eq:zeta} that we had before. Following the same derivation used in the proof of Theorem \ref{thrm:regular_GLDPC}, with the exception that $\eta$ is replaced by $\tilde{\eta}$, yields (in parallel to \eqref{eq:pe_alpha_bndf}),
\begin{equation}
\label{eq:pe_alpha_bndf_r}
p_e(\alpha N, \nu N)
\le
C_3 \cdot \left( \alpha N \right)^{11/2} e^{N \tilde{f}(\alpha,\nu)}
\end{equation}
\begin{equation}
\label{eq:falpha_r}
\tilde{f}(\alpha,\nu) \defined \max_{\gamma,\delta,\phi,\omega} \tilde{\psi}(\alpha,\gamma,\delta,\phi,\omega) \: .
\end{equation}
The maximization is over values of $\gamma, \delta, \phi$ and $\omega$ that satisfy~\eqref{eq:gamma_bnd}-\eqref{eq:gamma_bnd1} and $\gamma+\delta \ge \nu$. The function $\tilde{\psi}()$ is defined by,
\begin{equation}
\tilde{\psi}(\alpha,\gamma,\delta,\phi,\omega)
\defined
\alpha h(\gamma/\alpha) + (1-\alpha) h(\delta/(1-\alpha))
+ (c/d) h(\phi)
+
\rho - c h(\omega,\alpha-\omega,\phi-\omega)
\label{eq:psi_r}
\end{equation}
and $\rho$ is given by \eqref{eq:rho}, as before.
Combining \eqref{eq:pe_alpha_bndf_r}-\eqref{eq:psi_r} with Theorem \ref{thrm:regular_GLDPC} yields the following theorem.
\begin{theorem}
\label{thrm:regular_GLDPC_random}
Consider the ensemble of $(c, d)$-regular binary GLDPC codes with blocklength $N$. Each component code has blocklength $d$ and it can correct all error patterns of size $t>0$ or less.
Suppose that the parallel bit flipping bounded distance decoding algorithm is used to decode a randomly chosen code from this ensemble under the condition \eqref{eq:c1_cond}.
Let $\alpha_R$ denote the largest positive value of $\alpha$ for which $\tilde{f}(\alpha,\alpha_0) \le 0$ where $\alpha_0$ is defined in Theorem \ref{thrm:regular_GLDPC}.
Then, if we start the decoding with $\lfloor \alpha N \rfloor$ errors, whose bit locations are determined uniformly at random, and $\alpha<\alpha_R$, the decoding will be successful with probability that approaches one as $N\rightarrow \infty$.
\end{theorem}

\beginproof
By the discussion above, with probability that approaches one as $N\rightarrow \infty$, after the first iteration there will be at most $\lfloor \overline{\alpha}_0 N \rfloor$ errors for $\overline{\alpha}_0 < \alpha$, and by Theorem \ref{thrm:regular_GLDPC}, all these remaining errors will be corrected in the following iterations. The required result now follows by using the union bound.
\finproof

\subsection{Non-binary regular GLDPC codes} \label{sec:nonbinary}

The parallel bit flipping bounded distance decoding algorithm for GLDPC codes was defined in
Algorithm \ref{alg:BDD} both for binary and non-binary codes over $\GF(q)$. In that description, the non-binary case is differentiated from the binary one by using brackets.
It can be observed that the same analysis that was applied in the proof of Theorem 1 holds also for non-binary GLDPC codes. In particular, Fig. \ref{fig:sets} also applies to non-binary GLDPC codes: After the iteration, each $v\in\cB_g$ will become correct, and each $v\in\cG_g$ will stay correct. The only difference between binary and non-binary GLDPC codes is that in the non-binary case we impose the additional restriction $c_1>c/2$ (see note 3 after the introduction of the algorithm), e.g., we can set $c_1 = \lfloor c / 2 \rfloor + 1$. Under this restriction Theorems \ref{thrm:regular_GLDPC} and \ref{thrm:regular_GLDPC_random} are still valid.
In particular, by Theorem \ref{thrm:regular_GLDPC}, for $N$ sufficiently large, under the conditions \eqref{eq:c1_cond} and $c_1>c/2$, standard random sampling of a non-binary GLDPC code from the ensemble yields, with probability approaching one, a good code that can correct a constant fraction of worst case errors.
Hence, the code can correct a constant fraction of worst case errors for $c\in\{3,4\}$ and all $t\ge 2$, and for $c=5$ and all $t\ge 1$ (set $c_1=\lfloor c / 2 \rfloor + 1$). Comparing to binary GLDPC codes, there is a difference only for $c=4$: In this case, binary GLDPC codes require only $t\ge 1$, while non-binary GLDPC codes require $t\ge 2$.

\section{Numerical Examples} \label{sec:numerical_examples}
\begin{example}
\label{example:hamming_GLDPC}
Consider ensembles of $(c(m),d(m))$-regular binary GLDPC codes with an Hamming component code, $\cC_0(d(m),k_0(m))$, for $m = \{7,8,9,10,11\}$, where
\begin{align}
d(m) = 2^{m}-1 &= \{ 127, 255, 511, 1023, 2047 \}\\
k_0(m) = d(m)-m = 2^{m}-m-1 &= \{ 120, 247, 502, 1013, 2036 \} \: .
\end{align}
The rate of $\cC_0(d(m),k_0(m))$ is
\begin{equation}
R_0(m) = k_0(m) / d(m) = \{ 0.9449, 0.9686, 0.9824, 0.9902, 0.9946 \} \: .
\end{equation}
The minimum distance of $\cC_0(d(m),k_0(m))$ is $d_{\rm min}=3$ so that $t=1$. The left degree, $c(m)$, is set to make the nominal rate of the ensemble, $R(m)$, equal to about $1/2$. By \eqref{eq:R_GLDPC},
\begin{equation}
R(m) = 1 - c(m)(1-R_0(m)) \approx 1/2 \: .
\end{equation}
Hence, we set
\begin{equation}
c(m) = \{ 9, 16, 28, 51, 93 \} \: .
\end{equation}
The corresponding nominal rates of the ensembles are
\begin{equation}
R(m) = 1 - c(m)(1-R_0(m)) = \{ 0.5039, 0.4980, 0.5068, 0.5015, 0.5002 \} \: .
\end{equation}

This example was considered in \cite{zyablov2009low} under the parallel bit flipping bounded distance decoding algorithm. For each value of $m$, the maximum guaranteed fraction of worst case errors that can be corrected, denoted by `$\alpha_0$\cite{zyablov2009low}', for a sufficiently large blocklength, $N$, was computed by \cite[Theorem 1]{zyablov2009low} (note that `$\alpha_0$\cite{zyablov2009low}' is the same as $w_{\alpha}/2$ in \cite[Theorem 1]{zyablov2009low}). We then computed the maximum guaranteed fraction of worst case errors, $\alpha_0$, that can be corrected (for $N$ sufficiently large) by Theorem \ref{thrm:regular_GLDPC} (the smallest positive root of $f(\alpha)$ defined in \eqref{eq:falpha}). Finally, we computed the maximum guaranteed fraction of random errors, $\alpha_R$, that can be corrected (again, for $N$ sufficiently large) by Theorem \ref{thrm:regular_GLDPC_random}. 
In Fig. \ref{fig:err_frac_vs_d} we present a comparison between these bounds. As can be seen, for all values of $d=d(m)$, our guaranteed fraction of worst case errors that can be corrected, $\alpha_0$, is larger than the one obtained in \cite{zyablov2009low} ('$\alpha_0$\cite{zyablov2009low}'). The gap is larger for smaller values of $d$.
\begin{figure}[htbp]
\centering
\includegraphics[width=0.6\linewidth]{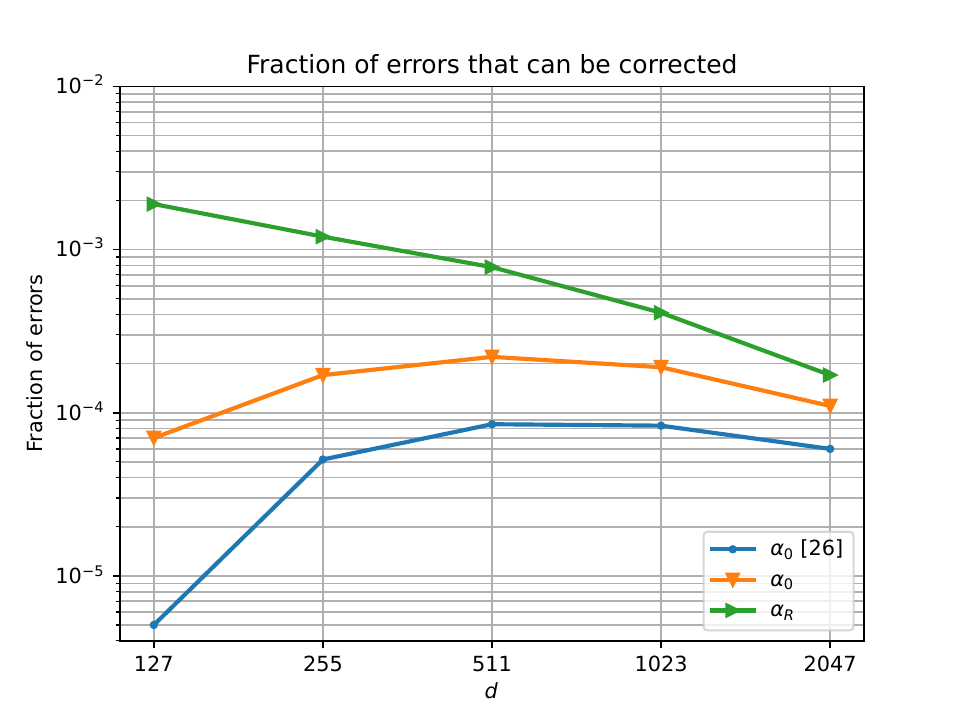} 
\caption{Maximum guaranteed fraction of errors that can be corrected vs. $d=d(m)$ for several Hamming GLDPC code ensembles. The curve denoted `$\alpha_0$ \cite{zyablov2009low}' presents the maximum guaranteed fraction of worst case errors that can be corrected according to \cite[Theorem 1]{zyablov2009low} (equal to $w_{\alpha}/2$ in \cite[Theorem 1]{zyablov2009low}). The curve  denoted `$\alpha_0$' ('$\alpha_R$', respectively) presents our maximum guaranteed fraction of worst case (random) errors that can be corrected according to Theorem \ref{thrm:regular_GLDPC} (Theorem \ref{thrm:regular_GLDPC_random}).
\label{fig:err_frac_vs_d}}
\end{figure}
The value of $\alpha_R$ is always larger than the corresponding $\alpha_0$ (and `$\alpha_0$\cite{zyablov2009low}') and the gap is larger for smaller values of $d$.
It can also be seen, for both $\alpha_0$ and `$\alpha_0$\cite{zyablov2009low}', that the largest guaranteed fraction of worst case errors that can be corrected is achieved for $d=511$. On the other hand, $\alpha_R$ is monotonically decreasing with $d$.

As was noted in \cite{zyablov2009low}, for all these code ensembles, the relative minimum distance of a typical code is very close to the Gilbert-Varshamov bound, which is about 0.11 (since the rate is about $1/2$). Hence, it is possible to correct a fraction of $0.055$ errors, while our bound, which applies to a simple decoder, guarantees much smaller fractions of correctable errors.

We have also examined the finite length behavior of our bound in this example for the code ensemble with $d=127$ (i.e., $m=7$).
An upper bound on the probability of failure to correct up to $\alpha_0 N$ worst case errors can be obtained by combining together \eqref{eq:pe}, \eqref{eq:pe_alpha}, \eqref{eq:eta_bnd_1} and \eqref{eq:zeta_bnd1a}. However, to further improve this bound we applied \eqref{eq:multi_bnd_2} (rather than \eqref{eq:multi_bnd_1}) to the two multiplying terms of $\eta$ in \eqref{eq:eta}.
To further improve the approximation to $p_e(\alpha N)$, we can replace the saddle point inequalities that lead to \eqref{eq:zeta_l_bnd} and \eqref{eq:zeta_r_bnd} by the Hayman method, e.g., \cite[Appendix D]{ru_book}. However, this would convert our bound to an approximation.
Fig. \ref{fig:err_prob} presents the resulting upper bound on the probability of failure to correct up to $\alpha_0 N$ worst case errors as a function of $\alpha_0 N$ for the code ensemble with $d=127$ and blocklength $N=\text{300,000}$.
\begin{figure}[htbp]
\centering
\includegraphics[width=0.6\linewidth]{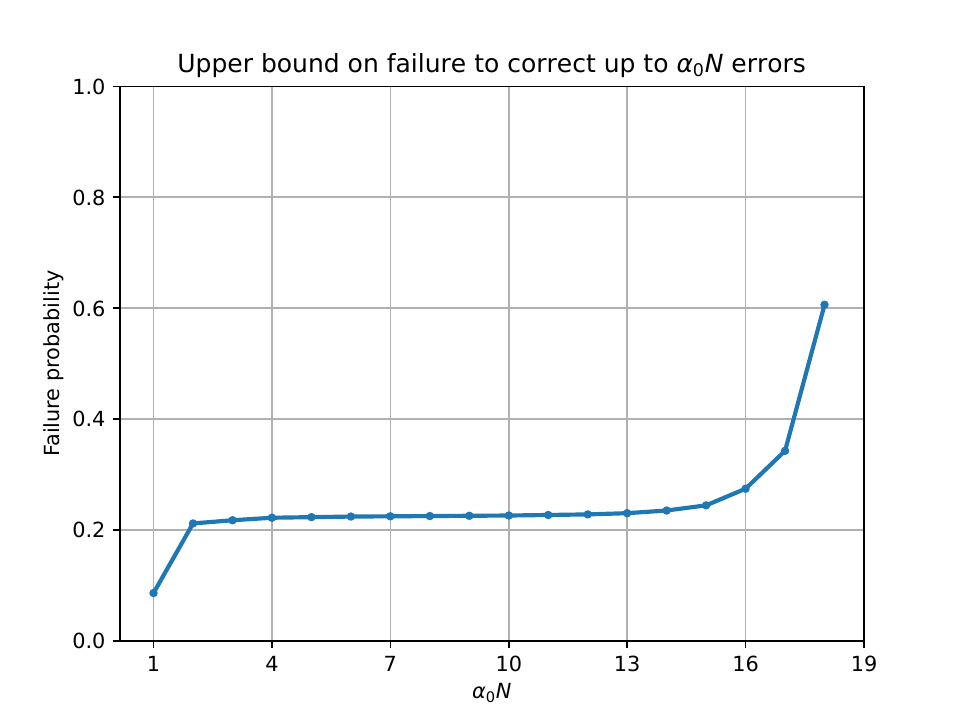} 
\caption{Upper bound on the probability of failure to correct up to $\alpha_0 N$ worst case errors as a function of $\alpha_0 N$ for the code with $d=127$. The blocklength is $N=\text{300,000}$.
\label{fig:err_prob}}
\end{figure}
The bound we plot is $\sum_{i=1}^{\alpha_0 N} \tilde{p}_e(i)$, where $\tilde{p}_e(i)$ is our upper bound on $p_e(i)$. Hence, the curve is monotonically increasing in $\alpha_0 N$. We plot the bound only for values of $\alpha_0 N$ for which $\sum_{i=1}^{\alpha_0 N} \tilde{p}_e(i) < 1$, since for larger values of $\alpha_0 N$ the bound is useless.
The figure shows that $\tilde{p}_e(1)$ and $\tilde{p}_e(2)$ are relatively large with significant contributions to the bound, while $\tilde{p}_e(i)$, for $i\in [3,13]$ are small. Then, for $i\ge 14$, $\tilde{p}_e(i)$ is a fast increasing function.
This behavior can be explained using the proof of Theorem \ref{thrm:regular_GLDPC}:
Combining \eqref{eq:pe_alpha_bndf} with \eqref{eq:noa_neglect}, that holds for sufficiently small $\alpha$, we see that for small $i$, $p_e(i)$ is upper bounded by a monotonically decreasing function (in $i$). On the other hand, for $\alpha_0$ defined in Theorem \ref{thrm:regular_GLDPC}, $f(\alpha_0)=0$. Hence, the terms $p_e(i)$ for $i$ close to $\alpha_0 N$ become larger.

It can be seen From Fig. \ref{fig:err_prob} that about $40\%$ of the codes in the ensemble can correct all error patterns of up to $18$ errors in any transmitted codeword when using the parallel bit flipping bounded distance decoding algorithm.

Once we have chosen a code at random from the ensemble, we can check if it contains a possibly bad  
$\Gamma(\alpha N, \gamma N, \delta N, \phi N c / d, \omega N c, c_1, t)$-partitioning for some small $\alpha N$ by an exhaustive search over the set $\cB$. If it does, then we discard the code and pick another one from the ensemble. In this way we can produce an expurgated code ensemble with bad codes (with respect to our decoding algorithm) eliminated. However, the exhaustive search is feasible only for very small values of $\alpha N$.
\end{example}

\begin{example}
\label{example:wc_errs}
Consider the ensemble of $(4,30)$-regular non-binary GLDPC codes with a component code, $\cC_0$, which is a Reed-Solomon (RS) code, RS($30,24$), and $q\ge 31$. This component code is maximum distance separable (MDS) and it satisfies the singleton bound with equality, i.e., its minimum distance is $d_{\min}=2t+1=30-24+1=7$, so that $t=3$.
The smallest positive root of $f(\alpha)$ defined in \eqref{eq:falpha} is $\alpha_0 = 0.0001$.
Hence, Theorem~\ref{thrm:regular_GLDPC} asserts, for a sufficiently large blocklength and a typical code in the ensemble, that Algorithm \ref{alg:BDD} can correct any fraction $\alpha\le 0.0001$ of errors in any transmitted codeword.

By \cite[Theorem 4]{chilappagari2010trapping}, Algorithm \ref{alg:BDD} can correct any fraction $\alpha\le \overline{\alpha}$ of errors in any transmitted codeword provided that the Tanner graph is a $(c,d,\overline{\alpha},\delta)$ expander for $\delta > (t+2)/(2(t+1))=0.625$.
Using \cite[Lemma 4]{bur_flipping} on the expansion of a random graph, we know that the above requirement on $\delta$ holds in a typical graph for $\overline{\alpha} = 7.3 \times 10^{-7}$ (this is an improvement compared to \cite{ru_book}[Theorem 8.7], by which $\overline{\alpha} = 1.7 \times 10^{-7}$).
Hence, compared to expander graph arguments, Theorem \ref{thrm:regular_GLDPC} asserts a much larger error correction radius (in addition to the fact that it relaxes the requirements on $t$ for fixed fraction worst case error correction for small $c$, as explained above).

Furthermore, for $\alpha_0=0.0001$, $\alpha_R$ defined in Theorem \ref{thrm:regular_GLDPC_random}, is given by $\alpha_R = 0.015$. Hence, when the blocklength is sufficiently large, Algorithm \ref{alg:BDD} can correct a fixed fraction of $0.015$ random errors in a typical code with probability that approaches one asymptotically in the blocklength.

The rate of $\cC_0$ is $R_0=0.8$. The nominal rate of the ensemble, as well as the rate of a typical code in the ensemble, is $1-(1-R_0)c=0.2$. For $q=31$, the Gilbert-Varshamov bound yields a relative minimum distance of about $0.611$. The corresponding fraction of errors that can be corrected with an appropriate decoder is about $0.3$.
\end{example}

\begin{example}
\label{example:wc_errs_c25}
Consider the ensemble of $(4,40)$-regular non-binary GLDPC codes with a component code, $\cC_0$, which is RS($40,32$), and $q\ge 41$. The minimum distance of $\cC_0$ is $d_{\min}=2t+1=40-32+1=9$, so that $t=4$.
In this case, Theorem~\ref{thrm:regular_GLDPC} asserts, for a sufficiently large blocklength and a typical code in the ensemble, that Algorithm \ref{alg:BDD} can correct any fraction $\alpha\le 0.00026$ of errors in any transmitted codeword.

By \cite[Theorem 4]{chilappagari2010trapping} combined with \cite[Lemma 4]{bur_flipping}, Algorithm \ref{alg:BDD} applied to a typical code in the ensemble, can correct any fraction $\alpha\le 2.4\times 10^{-6}$ of errors in any transmitted codeword (this is an improvement compared to \cite{ru_book}[Theorem 8.7], by which $\alpha \le 8.3 \times 10^{-7}$). Hence, once again, Theorem \ref{thrm:regular_GLDPC} asserts a much larger error correction radius.

Furthermore, for $\alpha_0=0.00026$, $\alpha_R$ defined in Theorem \ref{thrm:regular_GLDPC_random}, is given by $\alpha_R = 0.027$. Hence, when the blocklength is sufficiently large, Algorithm \ref{alg:BDD} can correct a fixed fraction of $0.027$ random errors in a typical code with probability that approaches one asymptotically.

The rate of $\cC_0$ is $R_0=0.8$. The nominal rate of the ensemble, as well as the rate of a typical code in the ensemble, is $1-(1-R_0)c=0.2$. For $q=41$, the Gilbert-Varshamov bound yields a relative minimum distance of about $0.626$. The corresponding fraction of errors that can be corrected with an appropriate decoder is about $0.31$.
\end{example}

\section{Conclusion} \label{sec:conclusion}
We have presented relaxed sufficient conditions under which a typical code in the regular GLDPC code ensemble, with blocklength sufficiently large, can correct a positive constant fraction of worst case errors.
We have also extended the analysis to random error correction and to non-binary GLDPC codes. The numerical examples demonstrate significant improvements on the fraction of worst case errors that can be corrected compared to the bounds predicted by expander graph arguments. These examples also demonstrate a much larger fraction of random errors that can be corrected compared to the fraction of correctable worst case errors.

To keep the analysis simple, we used the bit flipping bounded distance decoding algorithm. It might be possible to obtain improved results with a more complicated decoding algorithm, e.g., one that can also pass erasure messages.

In Section \ref{sec:numerical_examples}
we have suggested an expurgation scheme that eliminates bipartite graphs with possibly bad $\Gamma(\alpha N, \gamma N, \delta N, \phi N c / d, \omega N c, c_1, t)$-partitioning for small $\alpha N$. This scheme requires an exhaustive search and hence is suitable only for very small values of $\alpha N$. In practice, it may be possible to suggest methods for either detecting or avoiding possibly bad $\Gamma(\alpha N, \gamma N, \delta N, \phi N c / d, \omega N c, c_1, t)$-partitions for larger values of $\alpha N$. Note that this problem is similar to the detection or avoidance of small size trapping sets.





\end{document}